\documentclass[showpacs,prd]{revtex4}
\begin{document}
\newcommand{\Sw}{Schwarzschild }
\newcommand{\be}{\begin{equation}}
\newcommand{\ee}{\end{equation}}

\title{Entropy of Spacelike Two-Surfaces of Spacetime}

\author{Jarmo M\"akel\"a} 
\email[Electronic address: ]{jarmo.makela@phys.jyu.fi}  
\author{Ari Peltola}
\email[Electronic address: ]{ari.peltola@phys.jyu.fi} 
\affiliation{Department of Physics, University of Jyv\"askyl\"a, PB 35 (YFL), FIN-40351 Jyv\"askyl\"a, Finland}

\begin{abstract} We argue that every finite piecewise smooth spacelike two-surface of spacetime possesses an entropy which is, in natural units,
one quarter of its area. However, the thermodynamical properties of a two-surface become apparent only to the observers having that two-surface
as a horizon. Consequences of this result are discussed.
\end{abstract}


\maketitle
The concept of black hole entropy was first properly introduced by Bekenstein in 1972 \cite{bek}.
Inspired by Bekenstein's ideas, Hawking was able to show that black holes emit thermal radiation and that an exact expression for the black hole entropy is, 
in SI units,
\be \label{eq:entropy} S = \frac{1}{4}\frac{k_B c^3}{\hbar G}A,
\ee
where $A$ is the event horizon area of the black hole \cite{haw}. After
Bekenstein and Hawking had published their results it was found by Unruh that also the so-called Rindler horizon of an accelerated 
observer emits thermal radiation \cite{unruh}. This suggested that
Rindler horizon possesses entropy in the same way as black holes. Finally, it was shown by Padmanabhan that the usual thermodynamical 
properties, including entropy, may be associated with \textit{any} spacetime horizon, no matter whether we consider black hole, Rindler, de Sitter, or 
cosmological horizon \cite{pad}. In particular, it turns out that the entropy of a finite part of any horizon is, in natural units, 
one quarter of the area of that part. Therefore, it appears that entropy is a universal property of horizons. 

Bekenstein's original ideas were motivated mostly by the similarities between the second law of thermodynamics 
and Hawking's area theorem, which states that the area of the
event horizon of a black hole cannot decrease in any (classical) process. Bekenstein was also able to find an 
expression for the black hole entropy by using information theory. It is a well-known fact 
that the entropy of a system measures the lack of information about the actual inner structure of the system. 
Bekenstein's idea was to consider the minimum increase in the area of an event horizon
resulting from a disappearance of a particle into the black hole. In this situation the entropy of a black 
hole can be identified as the loss of information associated with the particle. In this 
sense, the entropy of a black hole may be viewed to describe the lost information of matter inside the black hole.

It should be safe to say that at this stage of research we are not sure what happens to the
information which is lost inside a black hole. However, it is possible to find certain arguments
which suggest that the information concerning the microstates of a black hole is somehow related to the degrees of freedom at the event horizon 
of the hole. Such arguments are provided, for instance, by the successes of loop quantum gravity \cite{can}: If an appropriate canonical quantization is 
performed to the classical 
spacetime outside the black hole, and the event horizon is treated as a boundary, one obtains a theory which yields certain surface states. These states 
live on the horizon and they can be identified as the microstates of the black hole. The entropy of the hole can be calculated by counting the microstates.
Thus, loop quantum gravity makes a claim that the entropy of a black hole is due to the microscopic structure of spacetime at the horizon.
There are good reasons to believe that a similar result holds for
other spacetime horizons as well.
For instance, it was showed by Jacobson, 
among other things, that the thermal radiation of Rindler horizon causes a change in the metric at the horizon
in such a way that the area of the considered part of the horizon shrinks \cite{jac}. 
His analysis suggests that not only the entropy of a black hole
but also the entropy of a Rindler horizon is related to the structure of spacetime at the horizon.
As it comes to the other horizons of spacetime, it seems that they also
possess an entropy whose origin might be the same as in the cases of black hole and 
Rindler horizons.

These considerations prompt us to state the key assumption of our paper:
We assume that the entropy of any horizon is a consequence from the degrees of freedom at the horizon or, more precisely, from the microscopic structure of
spacetime at the horizon. Although, at this stage, there is no comprehensive proof of this statement, this view is held, at least implicitly, 
by many physicists and it is closely related to the so-called holographic
hypothesis \cite{loput}.
Viewed in this way, it would be more appropriate to talk about the
entropy of the event horizon rather than the entropy of a black hole.

What makes the spacetime horizons distinct from the other spacelike two-surfaces of spacetime? More precisely, why should the notion of entropy 
to be associated with the spacetime horizons only? 
We stated above that there are reasons to believe that the entropy corresponding to spacetime horizons is due to the microscopic degrees of freedom of
spacetime at the horizon. We should, however, expect that \textit{any} spacelike two-surface has at least some kind of notion of entropy since the
microscopic structure of spacetime should be the same for all spacelike two-surfaces. 
Therefore, we draw a general conclusion: \textit{Any finite part of any  
spacelike two-surface has an entropy which is one quarter of its area}. 

Provided that one believes in the key assumption of our paper, this conclusion of ours follows from the results derived for the Rindler horizon. 
In order to avoid any pathological behaviour we restrict our analysis to piecewise smooth spacelike two-surfaces only. Our line of reasoning can be 
expressed in the following five steps:
\begin{itemize}
\item[(a)] The entropy associated with each finite part of a Rindler horizon is one quarter of its area.
\item[(b)] Each finite spacelike two-plane is, according to a certain accelerated observer, a part of a Rindler horizon.
\item[(c)] Therefore, the entropy associated with each finite spacelike two-plane is one quarter of its area.
\item[(d)] Any piecewise smooth spacelike two-surface can be viewed as a union of infinitesimal spacelike two-planes, each having an entropy equal to one
quarter of its area.
\item[(e)] Therefore, the entropy associated with any finite piecewise smooth spacelike two-surface is one quarter of its area.
\end{itemize}

Why do we not observe the entropy attributed to those spacelike two-surfaces which are not horizons? The answer to this question is simple: 
Entropy of a system cannot be observed directly. In order to find a value for the entropy of a system we must measure its temperature.
Moreover, according to the first law of thermodynamics,
\be \label{eq:1PS} dE =TdS,
\ee
the temperature of a system is associated with the \textit{changes} in its entropy and energy. More precisely, we observe a system to possess certain
temperature because, during the act of observation, the entropy and
the energy of the system change such that the change of energy is
proportional to the change of entropy. 
(Note that we do not have to include a term analogous to $pdV$ to 
Eq. (\ref{eq:1PS}) since we do not consider any exterior work to be done to our system.)
But this is exactly what happens when the two-surface under consideration is a horizon: 
Thermal radiation of a horizon leads to a change in the spacetime geometry at the horizon
in such a way that its area, and therefore entropy, changes. Therefore the thermodynamical properties of horizons can be observed
by measuring the temperature of their radiation.
However, there exists no such radiation mechanism for other spacelike two-surfaces and, in particular, there is no change in entropy or energy. 
Therefore, the entropy of these spacelike two-surfaces remains ``hidden'' and can be observed only by the observers having those two-surfaces as horizons.

In this paper we have proposed that every spacelike two-surface
possesses an entropy proportional to its area. 
This proposal does not 
contradict with known physics and it brings clarity to the concept of gravitational entropy: Entropy is due to the microscopic structure of spacetime, and 
this microscopic structure, and therefore entropy, is the same for all spacelike two-surfaces with the same area, no matter whether the two-surface under 
consideration is a horizon or not. For the real understanding of these matters,
however, one still has a long way to go since a completely satisfactory theory of the microscopic structure of spacetime is still lacking.

One may ask, and perhaps even with some justification, what
actually is the point of this paper? How can its results be used?
Indeed, if we really claim that spacelike two-surfaces have a certain
entropy even when they are not horizons, but in that case the
entropy cannot be observed, it may appear that we are coming
dangerously close in spirit to the medieval scholastic arguments about
how many angels can sit on a head of a pin. After all, physics is a
science which predicts observations.

Even if our proposal that all spacelike two-surfaces have a certain
entropy did not have direct observational consequences (a more
detailed analysis may well reveal such consequences), however, the
proposal may turn out a useful guiding principle when one constructs
microscopic theories of gravitational entropy. It is possible that an
assumption that the presence of a horizon is necessary for the
production of gravitational entropy is ultimately an unnecessary
complication. The calculations of gravitational entropy may turn out
much easier if one simply ignores an assumption that a horizon is
looked at, and instead employs a line of thought which would produce
the same expression for the entropy of all spacelike two-surfaces (such
a line of thought has been used, for instance, in Ref. \cite{mp1}, for
the calculation of the entropy of the \Sw black hole). It would be an
interesting challenge to perform a detailed analysis about whether an
assumption of the presence of a horizon is really essential in the existing derivations of the
Bekenstein-Hawking entropy law from a microscopic theory of gravity, or
is it just an unnecessary complication which may be ignored. If it
turns out that horizons are not really essential in these derivations,
our proposal may possibly be elevated to the status of a general
principle which should be reproduced, at least in the semiclassical
limit, by every plausible microscopic theory of gravity.

\acknowledgements
We thank Jorma Louko and Markku Lehto for their constructive criticism during the preparation of this paper.

\end{document}